\documentclass[aps,prb,twocolumn,groupedaddress,showpacs,reprint]{revtex4-2}
\usepackage{graphicx}
\usepackage{amsmath}
\usepackage{amssymb}
\usepackage{color}
\newcommand{\exciting}{{\usefont{T1}{lmtt}{b}{n}exciting}}

\begin{document}
	
	\title{Influence of spin-orbit coupling on chemical bonding }
	\author{Andris Gulans$^{1,2}$}
	\author{Claudia Draxl$^2$}
	\affiliation{$^1$Department of Physics, University of Latvia, Jelgavas iela 3, Riga, LV-1004 Latvia}
	\affiliation{$^2$Physics Department and IRIS Adlershof, Humboldt-Universit\"at zu Berlin, Zum Gro\ss en Windkanal 2, 12489 Berlin, Germany}
	\date{\today}
	\begin{abstract}
		The influence of spin-orbit interaction on chemical bonds in elemental solids and homonuclear dimers is analyzed by means of density-functional-theory calculations.
		Employing highly precise all-electron full-potential methodology, our results represent benchmark quality.
		Comparison of the scalar- and fully-relativistic approaches for elemental solids shows that the spin-orbit interaction may contract or expand the volume of the considered material.
		The largest variation of the volume is obtained for Au, Tl, I, Bi, Po and Hg, exhibiting changes between 1.0--7.6\%. 
		Using the tight-binding model, we show for diatomic molecules that the nature of this effect lies in the angular rearrangement of bonding and antibonding orbitals introduced by spin-orbit coupling.
		Such an angular rearrangement appears in partially filled $p$- or $d$-orbitals in heavy elements. 
		Finally, we discuss the impact of the relativistic effects on the chemical bonding in single-layer iodides and transition metal dichalcogenides. 
		
	\end{abstract}
	
	\maketitle
	
	\section{Introduction}
	
	Relativistic quantum theory prescribes that the spin of an electron is coupled with its spatial motion.
	This effect is known as the spin-orbit coupling (SOC), and it gives rise to a number of physical and chemical properties, 
	such as the band gap in graphene and topological insulating phases. 
	Previous studies have shown that SOC not only impacts electronic bands but also symmetry \cite{Pyykko2012} and chemical bonding, particularly in solids with heavy elements \cite{Legut2007,Hermann2010,Oliveira_2013,Sanchez2007}.
	For example, the volume of simple cubic Po expands by 5\% due to the spin-orbit interaction \cite{Legut2007} demonstrating that there are materials where this effect is sizeable and must be accounted for.
	
	The vast majority of calculations performed using Kohn-Sham density-functional theory (DFT) \cite{Hohenberg1964,Kohn1965} employ scalar-relativistic approximations, some even disregard relativity completely.
	Such a choice does not affect bond lengths in materials with light elements, but is questionable when it comes to materials containing heavy atoms.
	A few previous studies \cite{Philipsen2000,Hermann2010,Lejaeghere2014} have compared structural and elastic properties of elemental solids as calculated with the scalar- and fully relativistic approximations, where e.g. Refs.~\cite{Hermann2010} and \cite{Philipsen2000} were focusing on just a few specific groups of the Periodic Table of Elements (PTE). 
	Such selection exemplified a few chemical elements for which  SOC plays an important role but it does not allow one to make conclusions about the trends in the PTE. 
	In Ref.~\cite{Lejaeghere2014}, the complete 5$d$ and almost the complete 6$p$ series are discussed.
	However, the presented data are not provided in detail such to make them useful for quantitative analysis and, hence, benchmarking purposes.
	None of these three studies provide a detailed reasoning why and how  SOC impacts the bond lengths.
	Such a discussion is available only in exceptional cases, like for the expansion of the bond in Bi$_2$ \cite{Oliveira2013} using the tight-binding framework.
	Similarly, Ref.~\cite{Zeng2011} presents the mechanism how  SOC influences the chemical bond in Tl$_2$ and TlH.
	A more general discussion on the impact of  SOC on chemical bonds is still missing. 
	%\textbf{Is this statement too careless?}
	
	The computational materials-science community performs scalar-relativistic calculations on a daily basis, yet the numerical precision for capturing subtle effects may not always be taken for granted.
	A recent study~\cite{Lejaeghereaad3000} has shown that modern DFT codes show a sufficient level of reproducibility for elemental solids, but this outcome was a result of a concerted effort.
	Here, we go beyond by involving fully-relativistic calculations which are substantially more demanding.
	First, the spin-orbit interaction requires a spinor formalism, doubling the degrees of freedom.
	The frequently used second-variational approach~\cite{SinghD94} allows one to work with just a small overhead compared to the cost of scalar-relativistic calculations, but the numerical uncertainties related to this approximation are often difficult to control.
	Second, in contrast to scalar-relativistic theory, fully-relativistic wavefunctions exhibit a different asymptotic power law in the vicinity of the nuclei.  
	As a consequence, one distinguishes between states with the same angular momentum $l$, but different total  momentum $j$. Overall,  SOC effects mainly originate from the spatial region close to the nuclei, and this is where many DFT codes make use of approximations such as employing pseudopotentials.
	Having these two points in mind, it is natural to expect that including  SOC presents an additional challenge concerning precision and reproducibility of DFT data.
	
	In this paper, we present our treatment of relativity in the all-electron full-potential code \exciting{} \cite{Gulans2014a} which is suitable for reaching microhartree precision as demonstrated in Ref.~\onlinecite{Gulans2018} for light atoms in the non-relativistic case.
	%We show that our implementation is suitable for reaching microhartree precision, similarly to what was demonstrated in Ref.~\onlinecite{Gulans2018} for light atoms in the non-relativistic case.
	Such a feature is vitally important for the quality control of calculations and benchmarking, because only controlling the numerical precision enables systematic improvements what concerns the accuracy of the underlying methodology.
	We use this implementation for performing high-quality DFT calculations of the equation of state 
	employing scalar- and fully-relativistic theory, where we consider 55 non-magnetic elemental solids of the periods 4--6.
	We demonstrate the effect of  SOC on the lattice parameters and provide a qualitative explanation to it based on a tight-binding model applied to dimers of $p$-elements.
	Finally, we discuss this effect in two-dimensionsional transition-metal dichalcogenides (TMDCs) and group-IIB and -IVA iodides.

	\section{Method}
	\subsection{Relativity in DFT calculations}
	The relativistic Kohn-Sham theory is based on approximations to the four-component Dirac equation which, written for the large component $\Psi$ and the small component $\Phi$ in matrix form, reads  
	\begin{equation}
	\label{eq:dirac}
	\left( \begin{array}{cc}
	V & c(\boldsymbol{\sigma p}) \\
	c(\boldsymbol{\sigma p}) & V-2c^2
	\end{array} 
	\right)
	\left( \begin{array}{c}
	\Psi \\
	\Phi
	\end{array}
	\right) =
	E\left( \begin{array}{c}
	\Psi \\
	\Phi
	\end{array}
	\right),
	\end{equation}
	where $V$, $\mathbf{p}$, and $\boldsymbol{\sigma}$ are the effective Kohn-Sham potential, the momentum operator, and the vector of Pauli matrices, respectively.
	Formal elimination of the small component leads to an equation for the large component,
	\begin{equation}
	\label{eq:dirac-lc}
	\frac{1}{2}(\boldsymbol{\sigma p})\left[1+(E-V)/2c^2\right]^{-1}(\boldsymbol{\sigma p})\Psi+V\Psi=E\Psi.
	\end{equation}
	The factor $[1-(E-V)/2c^2]^{-1}$ in the kinetic-energy term introduces a non-trivial energy dependence which turns Eq.~\ref{eq:dirac-lc} into a non-linear eigenvalue problem.
	The simplest approach to make this equation tractable is to introduce the zero-order regular approximation (ZORA)~\cite{Lenthe1993,Lenthe1994}.
	Unfortunately, ZORA is not gauge invariant, namely, adding a constant to the effective potential $V$ results in a change of the total energy.
	Therefore, we apply the infinite-order relativistic approximation (IORA)~\cite{Dyall1999} that
	transforms Eq.~\ref{eq:dirac-lc} into
	\begin{equation}
	\label{eq:IORA}
	\begin{array}{l}
	\frac{1}{2}(\boldsymbol{\sigma p})\left(1-\frac{V}{2c^2}\right)^{-1}(\boldsymbol{\sigma p})\Psi+V\Psi = \\
	\frac{E}{4c^2}(\boldsymbol{\sigma p})\left(1-\frac{V}{2c^2}\right)^{-2}(\boldsymbol{\sigma p})\Psi+E\Psi.
	\end{array}
	\end{equation}
	
	This equation presents a generalized eigenvalue problem, and its solutions are the large components that are not orthogonal among themselves.
	Orthogonality is restored only for 4-spinors once the small component is considered too.
	In practice, however, we apply these relativistic approximations not to the deep-lying core states but to semicore, valence, and conduction bands, where the small component contributes to the density and the total energy only negligibly.
	Therefore, we restrict our calculations to considering only the large component that we normalize to 1.
	
	To proceed further with Eqs.~\ref{eq:dirac-lc}--\ref{eq:IORA}, one uses the following identity:
	\begin{equation}
	\label{eq:sr-so}
	\frac{1}{2}(\boldsymbol{\sigma p}) K (\boldsymbol{\sigma p})=\frac{1}{2}\mathbf{p}K\mathbf{p} + \frac{i}{2}\boldsymbol{\sigma}(\nabla K\times\mathbf{p}),
	\end{equation}
	where  $K$ is some function of $\mathbf{r}$, for example, $K=\left(1-\frac{V}{2c^2}\right)^{-1}$.
	The two terms on the right-hand side, $\frac{1}{2}\mathbf{p}K\mathbf{p}$ and $\frac{i}{2}\boldsymbol{\sigma}(\nabla K\times\mathbf{p})$, are known as the scalar-relativistic and spin-orbit contributions to the kinetic energy, respectively. Ignoring the latter, turns Eq.~\ref{eq:IORA} into the scalar-relativistic approximation:
	\begin{equation}
	\label{eq:SR-IORA}
	%\begin{array}{cc}
	\begin{split}
	\frac{1}{2}\boldsymbol{p}\left(1-\frac{V}{2c^2}\right)^{-1}\boldsymbol{p}\psi+V\psi = \\
	\frac{E}{4c^2}\boldsymbol{p}\left(1-\frac{V}{2c^2}\right)^{-2}\boldsymbol{p}\psi+E\psi.
	\end{split}
	%\end{array}
	\end{equation}
	
	\subsection{\label{sec:SR} Scalar-relativistic LAPW+LO}
	
	In this work, we employ the basis consisting of linearized augmented planewaves (LAPW) and local orbitals (LO). 
	For all other LAPW variants the procedure is analogous. 
	This method exploits our physical intuition regarding the shape of wave functions that 
	have atomic character in the vicinity of nuclei and vary slowly in regions far away from them.
	Therefore, the unit cell is partitioned into non-overlapping atomic spheres (\textit{MT$_{\alpha}$}, with $\alpha$ enumerating the atoms) and the interstitial region (\textit{I}).
	We use this division of space for defining LAPWs as atomic-like orbitals in the spheres and planewaves in the interstitial as:
	%
	%\begin{equation}
	%\label{eq:lapw-sr}
	%\phi_{\mathbf{G+k},\sigma}(\mathbf{r})=\sum\limits_{\xi{}lm} A^\mathbf{G+k}_{\xi{}lm} u_{\xi{}l}(r_\alpha; \varepsilon_{l}) Y_{lm}(\mathbf{\hat{r}_\alpha})\chi_\sigma,
	%  \frac{1}{\sqrt{\Omega}} e^{i(\mathbf{G+k})\mathbf{r}}\chi_\sigma & \mathbf{r}\in I
	%\end{equation}
	\begin{equation}
	\label{eq:lapw-sr}
	\begin{array}{ll}
	\phi_{\mathbf{G+k},\sigma}(\mathbf{r}) & =  A^\mathbf{G+k}_{lm} u_{l}(r_\alpha; \varepsilon_{l}) Y_{lm}(\mathbf{\hat{r}_\alpha})\chi_\sigma\\
	\\
	& + B^\mathbf{G+k}_{lm} \dot{u}_{l}(r_\alpha; \varepsilon_{l}) Y_{lm}(\mathbf{\hat{r}_\alpha})\chi_\sigma,
	\end{array}
	%  \frac{1}{\sqrt{\Omega}} e^{i(\mathbf{G+k})\mathbf{r}}\chi_\sigma & \mathbf{r}\in I
	\end{equation}
	and
	\begin{equation}
	\label{eq:lapw-sr-ir}
	\phi_{\mathbf{G+k},\sigma}(\mathbf{r})=\frac{1}{\sqrt{\Omega}} e^{i(\mathbf{G+k})\mathbf{r}}\chi_\sigma,
	\end{equation}
	respectively.
	The coefficients $A^\mathbf{G+k}_{lm}$ and $B^\mathbf{G+k}_{lm}$ are determined such that the LAPWs are continuous and smooth at the spheres boundaries.
	The radial parts of the atomic-like orbitals $u_l(r; \varepsilon_l)$ and their energy derivatives $\dot{u}_l(r; \varepsilon_l)$ are obtained from the scalar-relativistic radial Schr\"{o}dinger equation using a predefined energy parameter $\varepsilon_l$ for each $l$ value.
	
	Local orbitals (LO), the second type of basis functions, are defined as atomic-like functions that are non zero only in one atomic sphere:
	\begin{equation}
	\label{eq:lo-sr}
	\phi_{\mu,\sigma}(\mathbf{r})=  \left[a_{\mu} u_{l}(r_\alpha; \varepsilon_\mu) + b_{\mu} \tilde{u}_{l}(r_\alpha; \varepsilon_\mu^\prime)\right] Y_{lm}(\mathbf{\hat{r}})\chi_\sigma  .
	\end{equation}
	The coefficients $a_{\mu}$ and $b_{\mu}$ ensure that $\phi_{\mu,\sigma}(\mathbf{r})$ is normalized and turns to zero at the sphere boundary.
	The radial functions $u_{l}(r_\alpha; \varepsilon_\mu)$ and $\tilde{u}_{l}(r_\alpha; \varepsilon_\mu^\prime)$ are obtained from the scalar relativistic radial Schr\"{o}dinger equation similarly as above.
	Note that we do not restrict the definition of local orbitals just to radial functions and their derivatives ($u_l(r; \varepsilon_l)$, $\dot{u}_l(r; \varepsilon_l)$).
	Instead, $\tilde{u}$ can be chosen quite general, as an energy derivative of any order with a meaningful energy parameter.
	Such flexibility is important for achieving the basis-set limit.
	
	With the basis functions defined in Eqs.~\ref{eq:lapw-sr} and \ref{eq:lo-sr}, we now discuss the evaluation of matrix elements. 
	The non-relativistic case has already been described in detail in Ref.~\onlinecite{Gulans2014a}.
	The differences to the relativistic Hamiltonians are solely within the kinetic energy operator $\hat{T}$.  
	We evaluate the corresponding matrix elements consistently as
	\begin{equation}
	\label{eq:symmetricT-sr}
	T_{\nu\nu^\prime}^{\sigma\sigma^\prime}=\frac{1}{2} \delta_{\sigma\sigma^\prime} \int\limits_\Omega  [\boldsymbol{p} \phi^\sigma_\nu(\mathbf{r})]^\ast K [\boldsymbol{p} \phi^{\sigma^\prime}_{\nu^\prime}(\mathbf{r})] d^3r,
	\end{equation}
	where the integral is computed along similar lines as shown in Ref.~\cite{Gulans2014a}.
	
	\subsection{Fully-relativistic LAPW+LO \label{sec:FRLAPW}}
	
	The same basis, as defined in the previous section, can also be used in fully-relativistic calculations.
	However, such an approach is limited in accuracy, since the solutions of the atomic Dirac equation depend on the total angular momentum $j$.
	In particular, the radial parts of atomic wave functions obey different power laws depending on $j$ at small distances from a nucleus.
	Such a behavior has to be accounted for also in calculations going beyond atoms.
	The method described in Ref.~\onlinecite{Loucks1965} relies on relativistic augmented planewaves, where the basis fulfills the requirements of a fully-relativistic calculation.
	More specifically, the basis employs the radial Dirac equation for generating the radial parts of atomic-like orbitals (spinors) and combines them with the angular and spin degrees of freedom in terms of spin harmonics.
	
	In the present work, we rather employ a strategy that requires relatively small changes to the existing scalar-relativistic implementation of the LAPW+LO method.
	We keep the definitions of LAPWs and LOs as in Eqs.~\ref{eq:lapw-sr} and \ref{eq:lo-sr}, 
	except for the radial functions $u(r)$, $\dot{u}(r)$ and their further energy-derivatives that are now solutions of the radial Dirac equation. 
	Unlike in the previous section, the radial functions depend on the relativistic quantum number $\kappa$ 
	which is related to the angular and total angular momenta $l$ and $j$, respectively.
	In case of LAPWs, we set $\kappa=-(l+1)$ corresponding to $j=l+\frac{1}{2}$.
	In case of LOs, we choose $\kappa$ between the value above and $\kappa=l$ corresponding to $j=l-\frac{1}{2}$.
	A basis defined in such a way possesses all necessary degrees of freedom that are required for reliably solving the two-component problem in Eq.~\ref{eq:IORA}.
	
	Finally, we address another popular approximation. Typically, LAPW+LO implementations of the fully-relativistic approach consider  SOC only in the atomic spheres while ignoring the interstitial contributions~\cite{SinghD94}. Using the fact that the Kohn-Sham potential in the atomic spheres is approximately spherically symmetric, the expression of the kinetic energy operator in Eq.~\ref{eq:sr-so} transforms into 
	\begin{equation}
	\label{eq:sphere-so}
	\hat{T}= -\frac{1}{2}\boldsymbol{p}K\boldsymbol{p}+ \frac{K^2}{4c^2}\frac{1}{r}\frac{dV}{dr}\boldsymbol{\sigma L},
	\end{equation}
	where $\boldsymbol{L}$ is the angular momentum operator.
	We go beyond this approach, considering SOC also the interstitial region.
	Direct application of Eq.~\ref{eq:sr-so} leads to a non-Hermitian Hamiltonian in the LAPW+LO basis. 
	Instead, we calculate the matrix elements of the kinetic energy operator following the same spirit as in the scalar-relativistic case, {\it i.e.},
	\begin{equation}
	\label{eq:symmetricT}
	T_{\nu\nu^\prime}^{\sigma\sigma^\prime}=\frac{1}{2}\int\limits_\Omega  [\boldsymbol{\sigma p} \phi^\sigma_\nu(\mathbf{r})]^\ast K [\boldsymbol{\sigma p} \phi^{\sigma^\prime}_{\nu^\prime}(\mathbf{r})] d^3r,
	\end{equation}
	and, therefore, hermiticity of the Hamiltonian is preserved.
	
	Once the matrix elements are computed, we have to solve a full generalized eigenvalue problem that now corresponds to a fully-relativistic Hamiltonian.
	We note that the usual strategy for treating the spin-orbit coupling involves the second-variational approach, which is a two-step procedure. 
	First, one considers only the scalar-relativistic part of the Hamiltonian. 
	Then, a limited number of the so-obtained solutions is used as the basis for the full two-component problem~\cite{Koelling1974,MacDonald1980}.
	As it may be difficult to control the precision in the latter approach, we do not employ it but solve the full spinor problem.

	\section{Computational details \label{sec:details}}
	
	To quantify differences between scalar- and fully-relativistic calculations of elemental solids, we employ the protocol introduced in Refs.~\cite{Lejaeghere2014} and \cite{Lejaeghereaad3000}.
	In short, this protocol consists of the following steps: 
	(i) The unit cell of a chosen material is considered for 7 volumes ranging from 94\% to 106\% of the equilibrium volume $V_0$.
	(ii) The Birch-Murnaghan equation of state (EOS) is fitted to the obtained total energies and set to zero at the energy minimum.
	(iii) The EOSs, $E_1(V)$ and $E_2(V)$, obtained for a given solid with two different methods are compared. To this extent, the difference is quantified by the so-called $\Delta$-factor defined as
	\begin{equation}
	\label{eq:delta}
	\Delta=\sqrt{\frac{\int [E_1(V)-E_2(V)]^2 dV}{\Delta V} },
	\end{equation}
	where the integration is performed within the range of the volumes indicated above.
	
	The unit cells and atom positions are chosen like in Ref.~\cite{Lejaeghere2014}.
	Following Refs.~\cite{Lejaeghere2014} and \cite{Lejaeghereaad3000}, we remain within the generalized gradient approximation (GGA) as parametrized by Perdew, Burke, and Ernzernhof~\cite{Perdew1996} for exchange-correlation effects. The LAPW cutoff is set to $R_\mathrm{MT}G_\mathrm{max}=10$ and 12 for hydrogen and other chemical species, respectively.
	Such settings ensure highly converged EOSs as verified in Ref.~\cite{Gulansnomad2016}.
	Further computational details, such as the sets of LOs and $\mathbf{k}$-point grids, are available in Ref.~\cite{Gulansnomad2019}
	
	\begin{figure*}[bht]
		\includegraphics[width=17.8cm]{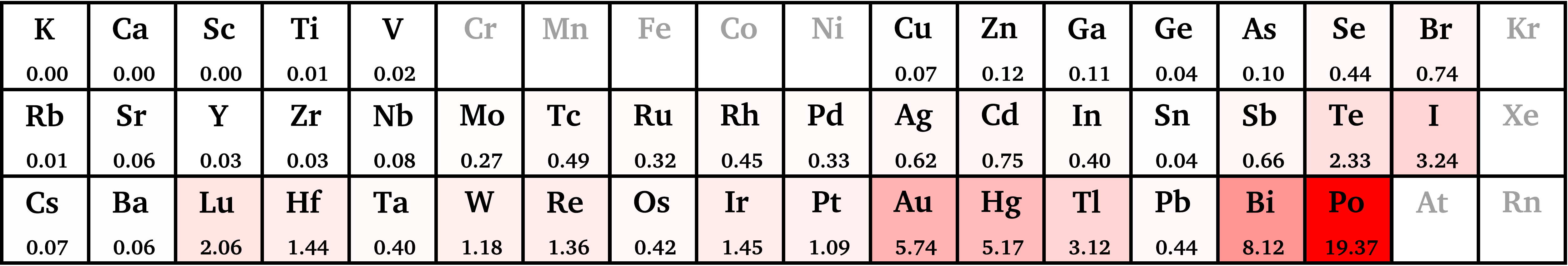}
		\caption{\label{fig:sodeltas} 
			$\Delta$ values (in meV/atom) for comparison between fully relativistic and scalar-relativistic calculations. 
			The intensity of the background color of each cell is proportional to $\Delta$ obtained for the respective element. 	} 
	\end{figure*}
	\begin{figure*}[htb]
		\includegraphics[width=17.8cm]{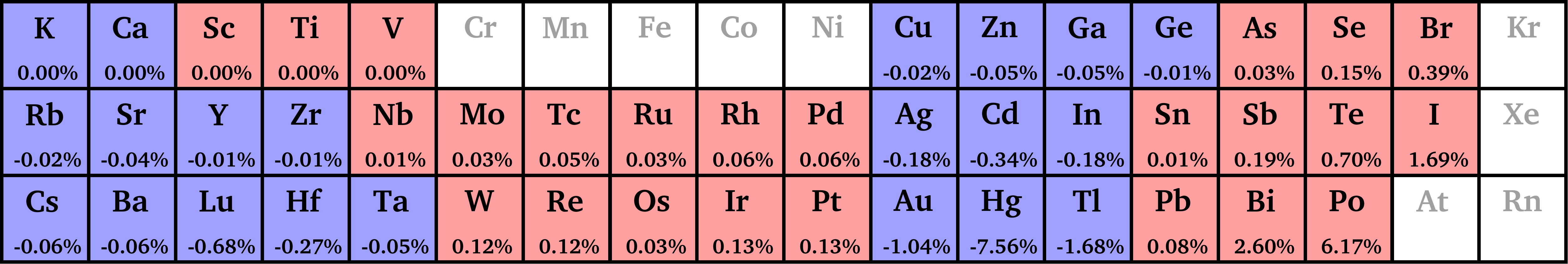}
		\caption{\label{fig:volumes} Relative change of volume in elemental solids due to spin-orbit coupling.
			Red and blue colors indicate expansion and contraction of the elemental solid, respectively.} 
	\end{figure*}
	
	Since SOC is expected to impact only heavy atoms, we do not consider the elements of the first three rows of the PTE. Furthermore, we exclude the magnetic fourth-period metals and the rare-gas solids. The latter are very poorly described by the PBE functional due to the lack of van der Waals interaction. In particular, 
	the weak spurious bonding obtained in PBE calculations is a consequence of the overlapping density tails of atoms.
	A fully-relativistic calculation of the magnetic fourth-period metals involves a non-collinear treatment, which goes beyond the scope of this paper.
	
	Calculations of dimers are performed in a unit cell with dimensions of $15\times 15\times 20$~bohr$^3$.
	The LAPW cutoff is set to $R_\mathrm{MT}G_\mathrm{max}=10$.
	
	Two-dimensional materials are modeled using the slab model with the thickness of the vacuum layer set to $\sim$10~\AA{}.
	The Brillouin zone is sampled by a grid of $9\times 9 \times 1$ $\mathbf{k}$-points.
	An LAPW cutoff of $R_\mathrm{MT}G_\mathrm{max}=12$ is used in these cases.

	\section{Results}
	\subsection{Elemental solids \label{sec:solids}}
	
	Figure ~\ref{fig:sodeltas} shows the impact of spin-orbit coupling in terms of $\Delta$ values, {\it i.e.}, differences between scalar- and fully-relativistic calculations.
	$\Delta$ ranges from 0.01~meV/atom (for K) to 19.37~meV/atom (for Po). 
	In other words, some elements are barely affected by SOC, while some of them experience a dramatic change.
	Overall, we observe two trends: 
	First, the effects increases with the period for elements within the same group.
	Second, within the same period, $\Delta$ does not increase monotonously with the nuclear charge, but rather follows a pattern.
	$\Delta$ is small for alkali and alkali-earth metals, and it reaches its maximum for chalcogens and halogens.
	Elemental solids of the other groups are affected by SOC to a variable degree, 
	and it is not straightforward to capture a pattern.
	However, we can already conclude that a few elements from the fifth period and more than half of those from the sixth period require a fully-relativistic treatment in order to obtain accurate EOSs.
	
	\begin{figure*}
		\includegraphics[width=17.8cm]{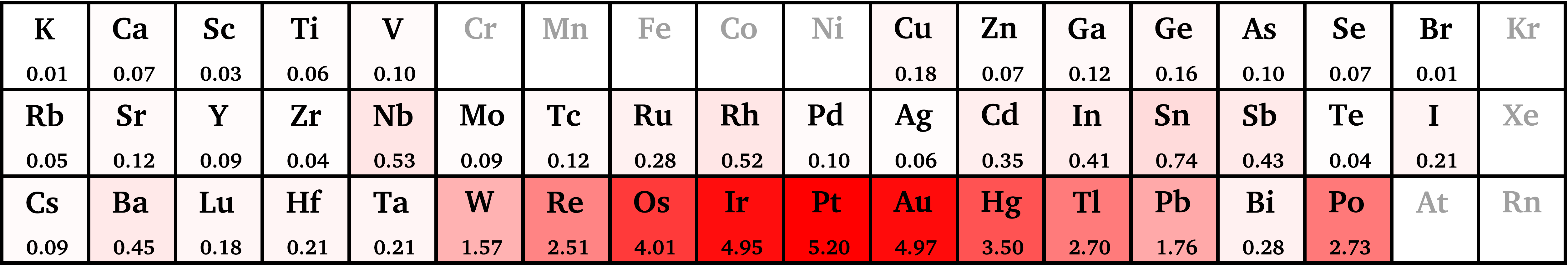}
		\caption{\label{fig:nojdeltas} $\Delta$ values (in meV/atom) evaluating fully-relativistic calculations performed with $j$-unresolved and $j$-resolved bases.
			The intensity of the color in the background of each cell scales proportionally to $\Delta$ obtained for each element.
		} 
	\end{figure*}
	
	\begin{figure}
		\includegraphics[width=8cm]{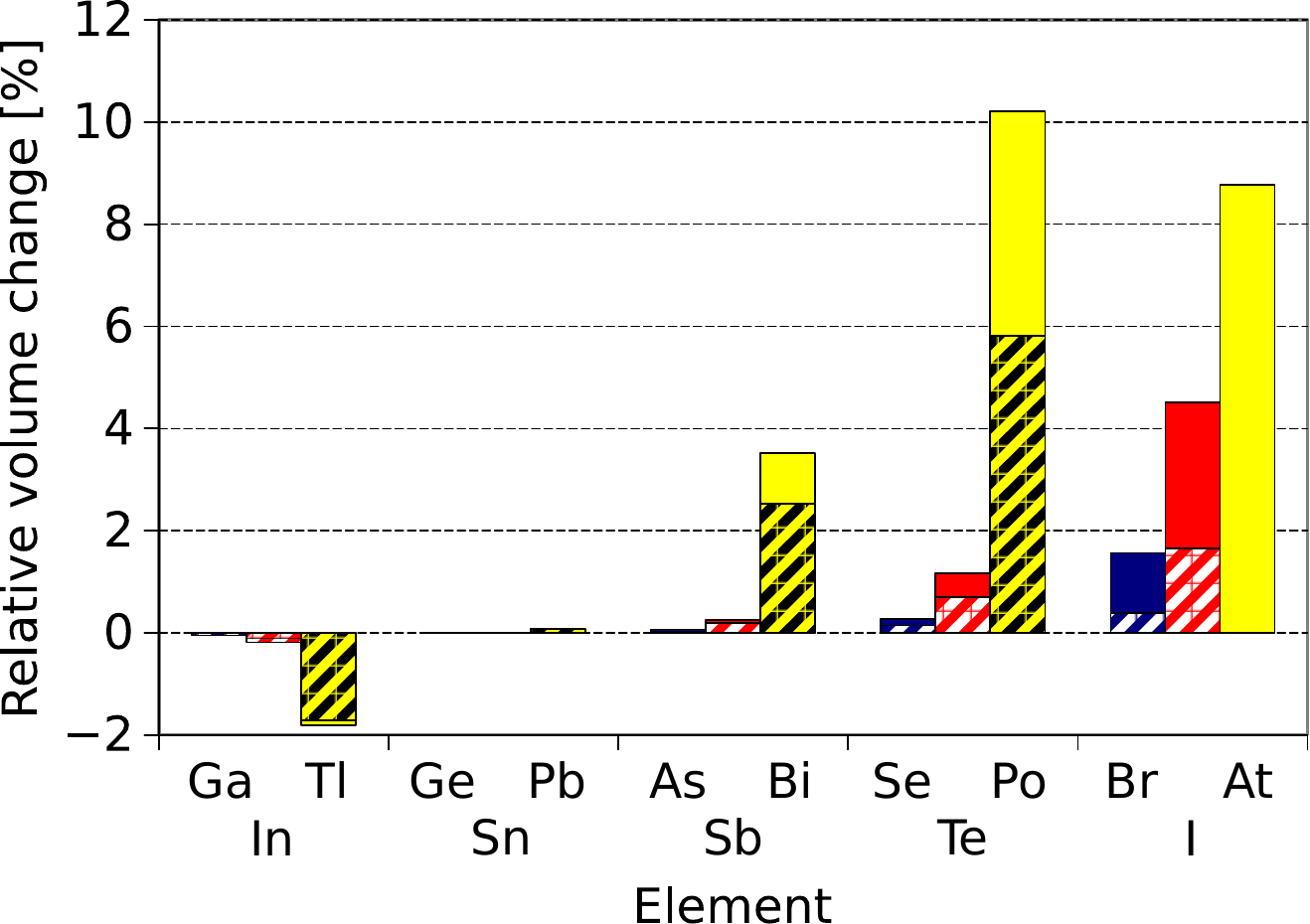}
		\caption{\label{fig:fcc-volumes} Relative change of volume in elemental solids due to spin-orbit coupling. 
			Full bars correspond to the fcc phase, whereas striped ones correspond to the structure given in Refs.~\cite{Lejaeghereaad3000} and \cite{Lejaeghere2014}. } 
	\end{figure}
	
	To get more insight into the observed trends, we analyze the calculated equilibrium volumes.
	In Fig.~\ref{fig:volumes}, their variation upon inclusion of SOC is illustrated.
	These results show that the equilibrium volume expands or shrinks following a strict pattern within the PTE.
	More specifically, we observe a volume reduction for elemental solids of groups IA--IIIA/VB and IB--IIIA/VIA and an expansion for the rest.
	The largest effects (in ascending order) are obtained for Au, Tl, I, Bi, Po and Hg, which either shrink or expand by $-7.6$--6.2\%.

	A number of previous studies have addressed this issue    \cite{Lejaeghere2014,Philipsen2000,Legut2007,Hermann2010}.
	Philipsen and Baerends have considered selected elemental solids, obtaining noticeable variations of the lattice constant in Au and Bi~\cite{Philipsen2000}.
	Similarly, Lejaeghere and coworkers \cite{Lejaeghere2014} have found that SOC has the strongest impact on the equilibrium volume of Re, Au, Po, Tl, Bi and Hg (in ascending order).
	This list contains five of the six elements highlighted by us but misses iodine that was not calculated fully relativistically in their work.
	A quantitative comparison with their results shows a major discrepancy for Po.
	According to Ref. \onlinecite{Lejaeghere2014}, its volumes is only changed by $\approx 1\%$, in stark contrast to our value of $6.1\%$.
	A different calculation of Po \cite{Legut2007} predicts a 5\%-volume increase, in much better agreement with our result.
	Finally, the volume contraction of Pb by as much as 8\% reported in Ref.~\onlinecite{Hermann2010} is in striking disagreement with the values of 0.1\% and 0.0\% obtained in this work and in  Ref. \onlinecite{Lejaeghere2014}, respectively.
	These contradictions highlight the importance of benchmark data and show that achieving high precision in fully-relativistic calculations appears to be still challenging.
	
	The above presented fully-relativistic data were obtained using the basis as described in Sec.~\ref{sec:FRLAPW}
	It is, however, quite common that in LAPW+LO calculations, in practice not a $j$-resolved basis is employed. In other words, the basis designed for scalar-relativistic calculations is often used also in fully-relativistic ones.
	Therefore, we explore the corresponding errors that are depicted in Fig.~\ref{fig:nojdeltas}.
	Our calculations show that this error is small for the elements of the fourth period.
	It then increases and becomes significant in the second half of the fifth period, with a maximum of 0.71~meV/atom obtained for Sn.
	An error of this magnitude, {\it i.e.}, the order of 1~meV/atom, is considered as a typical experimental uncertainty \cite{Lejaeghereaad3000}.
	Finally, in the sixth period, the error introduced by the $j$-unresolved basis increases even further and goes beyond the 1 meV/atom threshold for elemental solids from W to Pb as well as Po.
	The largest values of $\sim$5~meV/atom are reached for Ir, Pt and Au.
	In case of Au, the obtained $\Delta$ is comparable to the difference between the scalar- and fully-relativistic calculations reported in Fig.~\ref{fig:sodeltas}.
	Here the different basis sets lead to SOC-induced volume changes of about 0.2\% and 1.1\%, respectively.
	
	\begin{figure*}
		\includegraphics{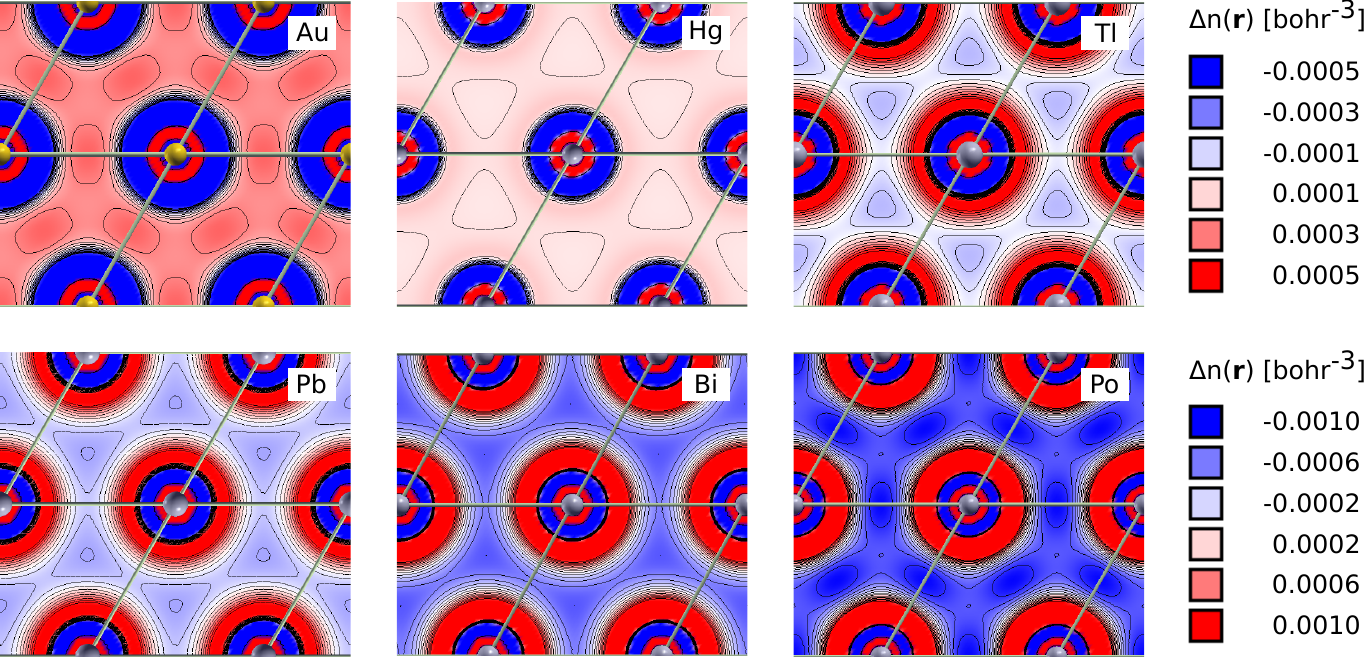}
		\caption{\label{fig:fcc-density} Rearrangement of electron density in fcc solids due to SOC in the (100) plane for Au, Hg, and Tl (top) as well as Pb, Bi, and Po (bottom).
			The contour lines correspond to 15 isovalues equally spaced between the minimum and maximum values of the respective scale where blue (red) indicates charge depletion (accumulation).
		} 
	\end{figure*}
	
	Since the considered elemental solids differ in terms of symmetry and atomic coordination, analyzing the trends across the PTE is intricate. To overcome this issue, we consider the $4p$-, $5p$- and $6p$-elements assuming the face-centered cubic (fcc) structure for all of them. Although there are numerical differences between the results for different lattice types, the qualitative picture is the same. For the SOC effect on their equilibrium volumes, we refer to Fig.~\ref{fig:fcc-volumes}. The group-IIIA elemental solids shrink, while those from the group IVA remain essentially unaffected. Finally, the elements from groups VA--VIIA expand, where the maximum expansion within each period is obtained for halogens, except for At in the sixth period.

	\subsection{Dimers}
	
	The results described in the previous section clearly show that SOC may introduce significant changes not only in the band structure, but also in the geometry of a solid.
	In other words, there are cases where determining the equilibrium structure in practice, requires a fully relativistic treatment.
	Here, we analyze why this is so and whether there is a way of telling in advance if it is necessary to include SOC.
	
	According to the data presented in Fig.~\ref{fig:volumes}, it increases the equilibrium volume of I by 1.7\%, which is among the largest volume expansions for the elemental solids. 
	Interestingly, as we will discuss below, this does not mean that the scalar-relativistic approximation is crude for all compounds containing this element.
	To show an example, we calculate the equilibrium volume of CsI in the rock-salt phase.
	The fully- and scalar-relativistic approaches yield values of 101.7~\AA$^3$ and 101.6~\AA$^3$ per formula unit, respectively.
	Thus, SOC expands the volume by only 0.1\%.
	Clearly, the difference between CsI and elemental I lies in the chemistry of the materials.
	
\begin{figure}
		\includegraphics[width=8.5cm]{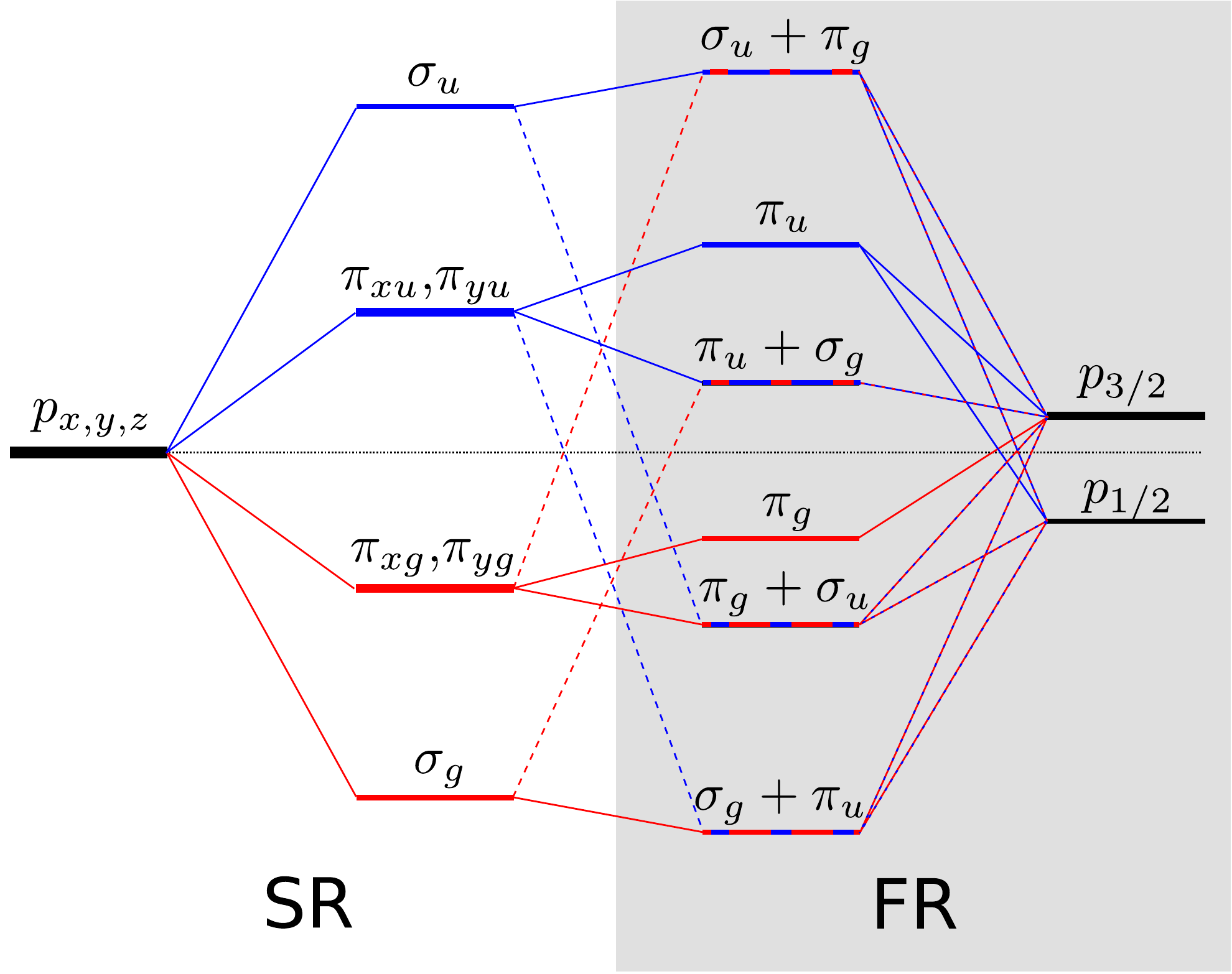}
		\caption{\label{fig:dimers-tb} Schematic of orbital energies in the I atom and the I$_2$ molecule as obtained from the tight-binding model.
			The labels SR and FR, correspond to the scalar- and fully-relativistic results, respectively.
			The two outer (inner) columns represent the atomic (molecular) energies.
			Red and blue lines correspond to bonding and anti-bonding orbitals, respectively. 
			For the states with mixed character both colors are used. 
		} 
	\end{figure}

	Bulk I is a is molecular crystal consisting of I$_2$ molecules, 
	and a volume change is basically attributed to a deformation of the I--I bond.
	To analyze the influence of SOC, we adopt the tight-binding model~\cite{Slater1954} with a basis that consists of atomic $5p$-orbitals. The corresponding Hamiltonian can be solved analytically in both the scalar- and fully-relativistic cases.
	Following the derivation presented in Ref.~\onlinecite{Oliveira2013}, we sketch a schematic of the orbital energies in Fig.~\ref{fig:dimers-tb}.
	The scalar-relativistic solution of the problem consists of three bonding ($\sigma_g$, $\pi_{xg}$,  $\pi_{yg}$) and three antibonding orbitals ($\sigma_u$, $\pi_{xu}$, $\pi_{yu}$).
	The $\pi_{xg}$ and $\pi_{yg}$ orbitals are degenerate, and so are $\pi_{xu}$ and $\pi_{yu}$.
	
	The fully-relativistic solution of the problem differentiates itself from the scalar-relativistic case in three qualitative points.
	First, the degeneracy of the $\pi$-orbitals is lost.
	Second, only two orbitals keep their angular character.
	The remaining four present a mixture of $\sigma$- and $\pi$-orbitals arising from the scalar-relativistic case, although in each case some particular character dominates.
	In Fig.~\ref{fig:dimers-tb}, this dominant contribution is indicated in the labels as the first one. 
	Third, only two orbitals keep their symmetry while the others do not have a pure bonding or antibonding character.
	Again, it is possible though to recognize in each case whether an orbital is predominantly bonding or antibonding.
	The second and third observations can be used for rationalizing the impact of SOC on the bond lengths. 
	As we show below, they can lead to contradictory conclusions in certain examples. 
	
	We start our discussion with the reasoning based on the population of bonding and antibonding orbitals~\cite{Oliveira2013}.
	An I$_2$ molecule in its ground state has 10 electrons in the $5p$-shells.
	They occupy the lowest-energy orbitals that include three bonding and two antibonding ones in the scalar-relativistic case.
	If SOC is considered, the only unoccupied level has $\sigma_\mathrm{u}$ character with a small contribution from $\pi_\mathrm{g}$.
	This means that electrons effectively occupy slightly less than three bonding and slightly more than two antibonding orbitals.
	Such a rearrangement weakens and thus elongates the chemical bond.
	We note in passing that these considerations not only apply to I$_2$ but also to other halogen dimers.
	
	To generalize this result, we consider dimers of a few other elements. Employing scalar-relativistic theory, dimers of the group-VA elements have fully occupied bonding and entirely empty antibonding orbitals.
	Like in the example above, inclusion of SOC changes this picture, and the effective population of the bonding orbitals reduces, thereby expanding the bonds.  
	
	%We also consider an alternative reasoning based on the angular momentum of electrons rather than the bonding character. 
	We also consider an alternative reasoning based on the redistribution of the angular character of electrons upon SOC.
	Our results sketched in Fig.~\ref{fig:dimers-tb} show that the fully-relativistic I$_2$ has more (fewer) electrons with $\sigma$ ($\pi$) character than the scalar-relativistic counterpart. As $\pi$ orbitals are more compact in the direction of the bond than $\sigma$ orbitals it is clear that the bond should expand upon spin-orbit interaction. 
	Applying the same analysis for group-VA dimers, the number of $\sigma$- and $\pi$-electrons changes by a small amount only. 
	The $\sigma_g$ state partially acquires $\pi$-character, forming a $\sigma_g+\pi_u$ state, 
	but one of the $\pi_g$ states partially loses $\pi$-character, forming a $\pi_g+\sigma_u$ state.
	Thus, this model predicts a small change of bond lengths in dimers such as Sb$_2$ and Bi$_2$.
	
	\begin{table}
		\caption{Bond lengths (in \AA) of homonuclear dimers in scalar (SR) and fully-relativistic (FR) calculations. 
			\label{tab:dimers}}
		%\begin{center}
		\setlength{\tabcolsep}{0.20cm} 
		\begin{tabular}{lcccccc}
			\hline
			\hline
			& As$_2$ & Sb$_2$ & Bi$_2$ & Br$_2$ & I$_2$ & At$_2$ \\
			\hline 
			SR & 2.110 & 2.476 & 2.651 & 2.309 & 2.684 & 2.868 \\
			FR & 2.110 & 2.477 & 2.678 & 2.310 & 2.704 & 3.036 \\
			\hline
			\hline
		\end{tabular}
		%\end{center}
	\end{table}
	
	To test the two qualitative arguments, we perform Kohn-Sham calculations of the group-V and group-VII dimers. The calculated bond lengths are given in Table~\ref{tab:dimers}. As predicted by the tight-binding calculations, the internuclear distances in the halogen molecules increase due to SOC. The variation is small (0.04\%) for Br$_2$, substantial (0.7\%) for I$_2$, and enormous (5.5\%) for At$_2$. SOC also increases the internuclear distances in the group-VA dimers, but just slightly in comparison to the halogens of the same periods. This small increase is in agreement with the argument based on the analysis of the $\sigma$/$\pi$ population but contradicts the explanation in terms of the (anti)bonding character of the orbitals.
	
	Although the presented tight-binding model is, strictly speaking, not justified for the group IIIA dimers, because of the triplet character of their ground state \cite{Lee2004,Zhou2013}, we nevertheless apply it to facilitate the discussion. 
	We argue that the chemical bond is expected to shorten, since it partially loses its $\sigma$ character and acquires some $\pi$ character in a fully-relativistic formalism; and this is what we indeed observe. 
	Thus, overall, depending on the filling of the $p$-shells, SOC results in shrinking or expanding chemical bonds.
	
	%However, the $\sigma$-$\pi$ splitting is so small in these elements that a sufficiently large spin-orbit splitting reverses the energetic order of the $\sigma_g+\pi_u$ and $\pi_g+\sigma_u$ orbitals.
	%This effect poses a challenge to a PBE calculation, where the self-consistent procedure reveals oscillations between the two solutions. 
	
	%This problem can be solved by introducing fractional occupations of orbitals or enforcing full occupation of either the $\sigma_g+\pi_u$ or the $\pi_g+\sigma_u$ orbitals.
	%Both options are merely technical workarounds, since a proper solution requires a multi-reference method~\cite{Tan2003,Lee2004,Zhou2013}. 
	%Furthermore, Refs.~\onlinecite{Lee2004,Zhou2013} report a triplet ground state that requires a different model than described in this section.

	\begin{figure}
		\includegraphics[width=8cm]{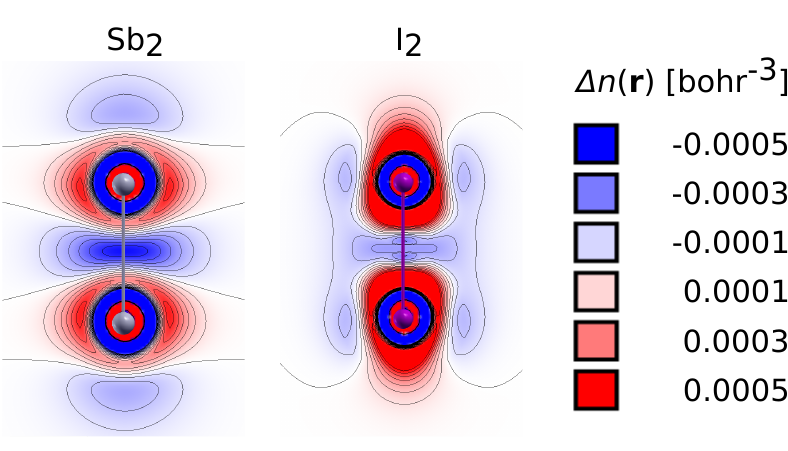}
		\caption{\label{fig:dimers-density} Rearrangement of electron density in homonuclear dimers due to spin-orbit interaction. Color maps are shown for the planes through the dimers. 
			The contour lines correspond to 15 isovalues equally spaced between the minimum and maximum values of the respective scale where blue (red) indicates charge depletion (accumulation).
		} 	
	\end{figure}
	
	The impact of SOC on the bonding in dimers and solids shows similar qualitative trends. Our results indicate that the key to a strong effect is the partially occupied $p$-shell. If this shell is completely filled as in the example of CsI above, SOC does not rearrange the angular character of the chemical bond, and therefore the volume remains essentially unchanged. Partially occupied $d$-shells play a role in the expansion or contraction of solids with $d$-elements. Almost empty and almost full $d$-shells lead to a contraction of a crystal, whereas half-filled $d$-shells give rise to an expansion.

	\section{2D materials}
	
	\begin{table}
		\caption{Calculated lattice constants, $a$, and bond lengths, $d$, (in \AA), and Bader charges, $q$, (in $e$) of metal ions in the investigated 2D materials from scalar- (SR) and fully-relativistic (FR) calculations. 
			\label{tab:2d}}
		\setlength{\tabcolsep}{0.18cm} 
		\begin{tabular}{lcccccc}
			\hline
			\hline
			& $a_\mathrm{SR}$ & $a_\mathrm{FR}$ & $d_\mathrm{SR}$ & $d_\mathrm{FR}$ & $q_\mathrm{SR}$ & $q_\mathrm{FR}$\\ 
			\hline
			MoS$_2$ & 3.1858 & 3.1858 & 2.4145 & 2.4145 & 1.215  & 1.215 \\
			MoSe$_2$ & 3.3200 & 3.3202 & 2.5419 & 2.5421 & 0.962 & 0.962 \\
			MoTe$_2$ & 3.5550 & 3.5565 & 2.7355 & 2.7363 & 0.591 & 0.590 \\
			WS$_2$ & 3.1881 & 3.1883 & 2.4204 & 2.4213 & 1.355 & 1.353 \\
			WSe$_2$ & 3.3209 & 3.3204 & 2.5486 & 2.5487 & 1.070 & 1.066 \\
			WTe$_2$ & 3.5587 & 3.5595 & 2.7410 & 2.7419 & 0.642 & 0.635 \\
			ZnI$_2$ & 4.0986 & 4.1058 & 2.8722 & 2.8761 & 0.787 & 0.776 \\
			CdI$_2$ & 4.3284 & 4.3342 & 3.0334 & 3.0366 & 0.793 & 0.780 \\
			HgI$_2$ & 4.3777 & 4.3795 & 3.0678 & 3.0667 & 0.451 & 0.414 \\
			GeI$_2$ & 4.3008 & 4.3013 & 3.0343 & 3.0367 & 0.727 & 0.716 \\
			SnI$_2$ & 4.5785 & 4.5743 & 3.2228 & 3.2251 & 0.924 & 0.907 \\
			PbI$_2$ & 4.6714 & 4.6527 & 3.2768 & 3.2778 & 0.947 & 0.879 \\
			\hline
			\hline
		\end{tabular}
		%\end{center}
	\end{table}
	
	The impact of SOC on chemical bonds goes beyond elemental solids. Here, we discuss this effect in single-layer 2D materials. Previous studies showed that many of them exhibit strong SO effects in the electronic structure impacting their response to light.
	It is natural then to ask whether SOC also has an impact on their structure.
	
	We focus on two classes of 2D materials, {\it i.e.}, iodides (AI$_2$, where A = Zn, Cd, Hg, Ge, Sn, Pb) in the PbI$_2$ structure and TMDCs (MX$_2$, where M = Mo, W and X = S, Se, Te) in the MoS$_2$ structure. 
	These geometries were observed experimentally in all cases except for HgI$_2$ \cite{Jeffrey1967}, for which we nevertheless use the PbI$_2$ lattice for the sake of consistency.
	
	\begin{figure}
		\includegraphics[width=8.5cm]{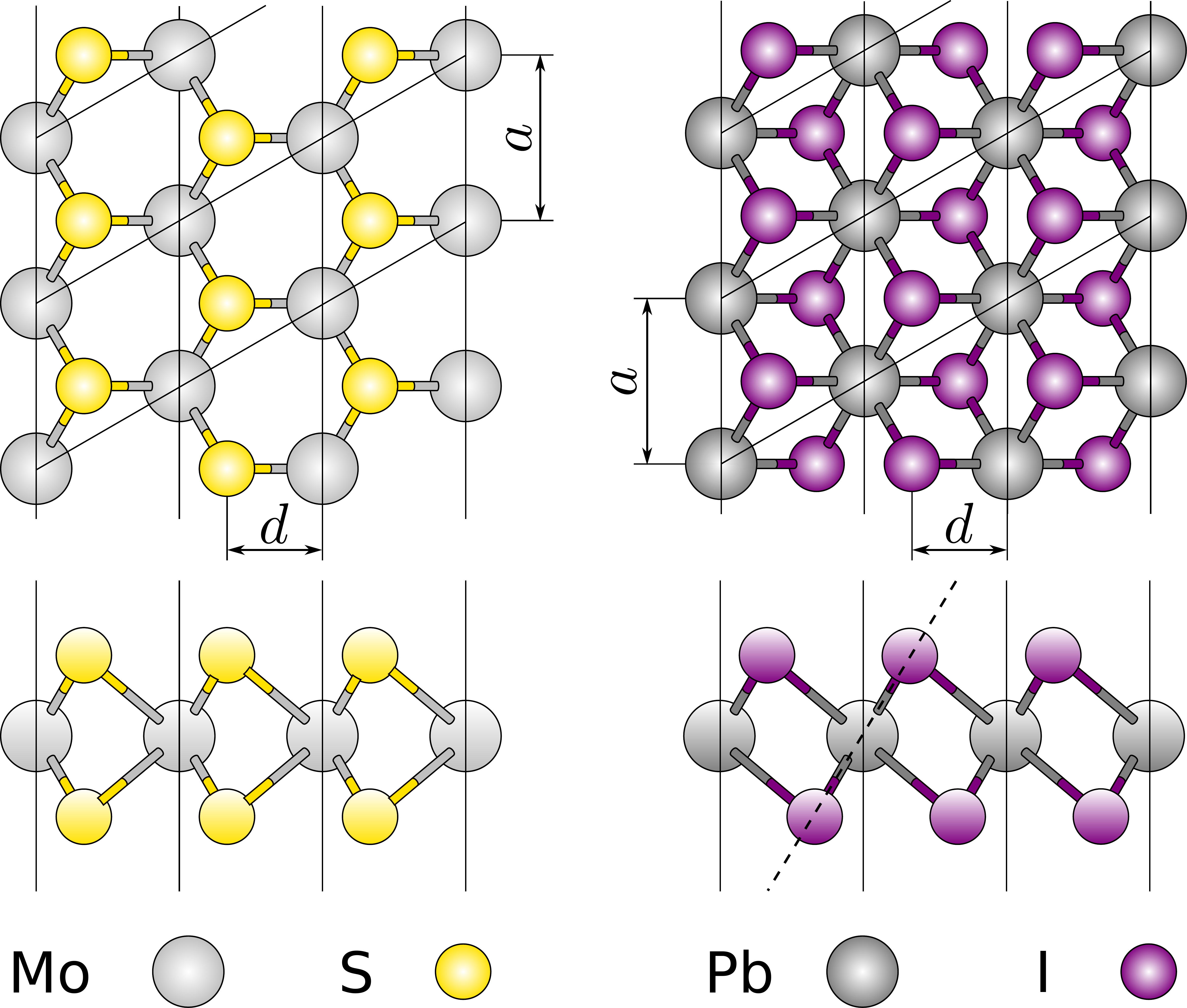}
		\caption{\label{fig:2d} 
			Structures of TMDCs and iodides shown with the examples of MoS$_2$ (left) and PbI$_2$ (right), respectively in top (top panels) and side views (bottom panels), respectively. 
			Thin full lines mark the boundaries of unit cells; the dashed line indicates the plane in which electron-density differences are shown in Fig.~\ref{fig:zni2-etc}. }
	\end{figure}
	
	Both considered structures are depicted in Fig.~\ref{fig:2d}. Owing to the high symmetry, the crystal lattice of the materials is uniquely defined by two parameters, {\it i.e.}, the lattice constant $a$, and the bond length $d$. We relax the structures with respect to both parameters. The results displayed in Table~\ref{tab:2d} show that the effect of SOC on the structure is generally weak for all considered TMDCs. The largest differences in $a$ and $d$ are observed for MoTe$_2$, where the increase of both is still less than 0.002~\AA. 
	We perform a Bader charge analysis \cite{Bader1990,Yu2011} which confirms that these materials are predominantly covalently bound with an admixture of ionicity, what is consistent with Ref.~\cite{Yue2013}. The charge of the metal ions ranges between 0.6~$e$ and 1.4~$e$. 
	Since the difference between the scalar- and fully-relativistic results is below 0.01~$e$ we conclude that  SOC does not impact the nature of bonding in these materials. 
	
\begin{figure*}
		\includegraphics[width=16cm]{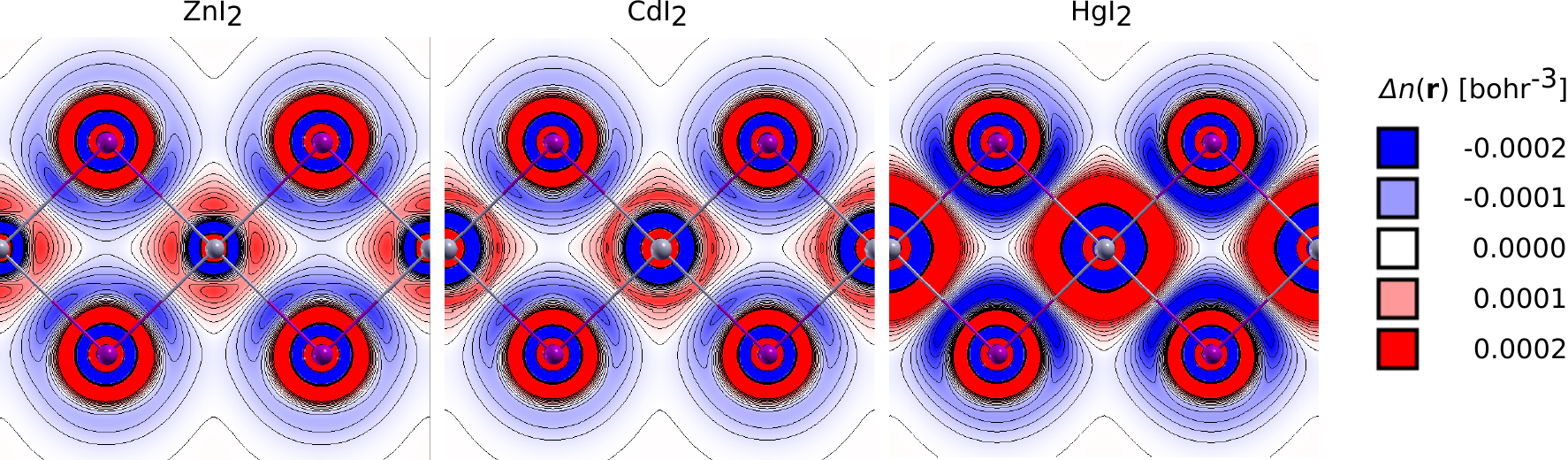}
		\caption{\label{fig:zni2-etc} 
			Rearrangement of the electron density in two-dimensional solids due to the spin-orbit interaction.
			The color maps are shown for the bonding planes passing through the I ion and cations. 
			The contour lines correspond to 15 isovalues equally spaced between the minimum and maximum values of the scale.} 
	\end{figure*}

	For the iodides, our calculations reveal a stronger effect of SOC. In particular, we find increased lattice constants in group-IIB iodides. 
	Remarkably, it is the largest for ZnI$_2$ despite Zn being the lightest cation of the group. This trend indicates that the lattice expansion in ZnI$_2$ is induced by relativistic effects on the I side. Replacing Zn by Cd or Hg, their presence counteracts the behavior of I, thus the chemical bonds are less affected. To illustrate this argument, we analyze the charge densities. SOC leads to a decrease of Bader charges of the metal ions by 0.012--0.035~$e$, making the materials slightly less ionic. This effect is smallest in ZnI$_2$, and it increases for the heavier group-IIB elements. To gain further insight, we plot charge-density differences, $\Delta q$, in (Fig.~\ref{fig:zni2-etc}). Like in the Bader charge analysis, we see that electrons are transferred from the I ions to the metal ions, with the following trend: $\Delta q_\mathrm{Zn}<\Delta q_\mathrm{Cd}<\Delta q_\mathrm{Hg}$. The transfer is accompanied by a rearrangement of the charge density between orbitals of different angular momentum around the nuclei. The strongest charge depletion happens right between the metal and the iodine nuclei. The largest increase, in turn, is visible in areas away from the direction of bonds.
	
	The lattice constant in group-IVA iodides reduces as SOC is taken into account. 
	This reduction is negligibly small in GeI$_2$, is larger in SnI$_2$, and reaches the maximum value of $\Delta a_\mathrm{SO}=0.0187$~\AA{} for PbI$_2$.
	Curiously, the Pb--I distance increases by 0.0010~\AA{}, while the reduction in $a$ is related to the change in the bond angle.
	Similarly to the other layered materials considered here, SOC makes also group-IVA iodides less ionic. 
	Also this effect is most pronounced in PbI$_2$ where the Bader charge of the Pb ions reduces by 0.066~$e$.
	
	The results presented in this section overall show that the SOC effect on the geometries is negligibly in most cases.
	The lattice constant changes by 0.02~\AA{} only in PbI$_2$, a change much smaller than, e.g., a difference caused by using PBE and LDA exchange-correlation functionals.
	However, we expect that the discussed relativistic effect may have a larger impact for the same materials with defects. 
	
	\section{Conclusions}
	We have presented and applied an implementation of fully-relativistic DFT within the framework of the linearized augmented planewaves plus local orbitals method. Our approach tackles the two-spinor problem directly without applying the second-variational approximation. To ensure high precision, we construct the basis by taking into account the $j$ dependence of the radial functions for all angular momenta corresponding to semicore and valence shells. 
	
	We have applied this method for calculating the equation-of-state parameters for elemental solids of the rows 4--6 of the PTE. The obtained results show that spin-orbit interaction influences equilibrium volumes and bulk moduli by modifying the chemical bonds. The strongest impact is observed for heavy elements with nearly empty or nearly complete $p$- or $d$-shells. Based on a tight-binding model, we trace this effect back to the angular redistribution of density in occupied bonding and anti-bonding orbitals.
	This result, obtained for elemental solids and homonuclear molecules, allows one to anticipate scenarios where SOC can impact the geometry of more complex materials in terms of chemical composition. We have illustrated this conclusion with the example of single-layer iodides. For the latter, we have found that the charge redistribution due to SOC introduces slight changes in the nature of bonding, making these materials less ionic. The largest change in the lattice constant among considered materials is about $0.02$~\AA{} obtained for PbI$_2$.

	\begin{acknowledgments}
		The work has received partial support from the European Union's Horizon 2020 research and innovation programme, grant agreement No. 676580 through the Center of Excellence NOMAD (Novel Materials Discovery Laboratory) and grant agreement No. 951786 through the Center of Excellence NOMAD (NOMAD CoE). We acknowledge also support from the German Research Foundation via SFB951, project number 182087777.
	\end{acknowledgments}
	
	%\bibliography{spin-orbit}

	%
	
\end{document}